\documentclass[12pt]{iopart}
\usepackage[dvips]{graphicx}
\begin{document}
\title{Investigation of the spin density wave in Na$_x$CoO$_2$}
\author{J Wooldridge, D McK Paul, G Balakrishnan and M R Lees}
\address{Department of Physics, University of Warwick, Coventry, CV4 7AL, United Kingdom}
\ead{J.Wooldridge@warwick.ac.uk}
\begin{abstract}
Magnetic susceptibility, transport and heat capacity measurements of single crystal $Na_xCoO_2$ (x=0.71) are reported. A transition to a spin density wave (SDW) state at T$_{mag}$ = 22 K is observable in all measurements, except $\rm \chi_{ac}$ data in which a cusp is observed at 4 K and attributed to a low temperature glassy phase. M(H) loops are hysteretic below 15 K. Both the SDW transition and low temperature hysteresis are only visible along the \textit{c}-axis. The system also exhibits a substantial ($\sim$40\%) positive magnetoresistance below this temperature. Calculations of the electronic heat capacity $\gamma$ above and below T$_{mag}$ and the size of the jump in C indicate that the onset of the SDW brings about the opening of a gap and the removal of part of the Fermi surface. Reduced in-plane electron-electron scattering counteracts the loss of carriers below the transition and as a result we see a net reduction in resistivity below T$_{mag}$. Sodium ordering transitions at higher temperatures are observable as peaks in the heat capacity with a corresponding increase in resistivity.
\end{abstract}
\submitto{\JPCM}
\maketitle

\section{Introduction}
$\rm Na_xCoO_2$ displays many interesting characteristics including unusual thermoelectric properties~\cite{wang}, charge and spin ordering~\cite{foo} as well as strong electron correlations~\cite{ando}. The structure is highly anisotropic; 2D layers of edge-sharing CoO$_6$ octahedra assembled incoherently along the \textit{c}-axis are separated by layers of partially occupied Na sites. The oxygen co-ordination of the Co sites results in a formal oxidation state of Co$^{3+}$ for NaCoO$_2$. A reduction in sodium corresponds to a doping of holes into the CoO$_2$ layer and consequently an increase in Co$^{4+}$ ions that have a low spin state. The Co d bands are crystal field split into a low lying ${\textit{t}}{_\textit{2g}}$ and an upper ${\textit{e}}{_\textit{g}}$ manifold. The ${\textit{t}}{_\textit{2g}}$ manifold is further split by the rhombohedral distortion within the $\rm CoO_6$ octahedra into two ${\textit{a}}{_\textit{1g}}$ and four ${\textit{e}}{\textit{'}}{_\textit{g}}$ bands. The Co ions sit on a two-dimensional triangular lattice which gives rise to magnetic frustration and this has prompted the suggestion of a resonance valence bond state~\cite{anderson, baskaran} as seen in the lanthanum cuprates, for example. Electron hopping on a triangular lattice can induce ferromagnetic correlations for the higher sodium concentrations; a weak ordered moment has been observed below T$_{mag}$ = 22 K by $\mu$SR studies~\cite{sugiyama}. The electron (and hence spin) hopping mechanism is thought to be the origin of the high thermopower observed in this system.
\par
The system has a heterogeneous phase diagram. Below one half stoichiometry (x$\leq \frac{1}{2}$) $\rm Na_xCoO_2$ is Pauli paramagnetic and it is from this range of sodium content that the newly discovered superconductivity in the hydrated compound originates~\cite{takada}. At $x=\frac{1}{2}$ an insulating phase is generated by sodium ordering on account of the localisation of charge carriers in the CoO$_2$ layers that experience a strong interaction with the charge density modulations in the sodium layer. Above half-doping, the system exhibits Curie-Weiss paramagnetism with a significant van Vleck term and the onset of weak magnetic ordering at T$_{mag}$ manifests itself as a spin density wave (SDW) that arrays antiferromagnetically perpendicular to the CoO$_2$ planes~\cite{sugiyama3}. 
\par
The recent interest in this system, in particularly for x$\geq$0.5, has generated many publications with some conflicting results. It is clear that studies on well-characterised, homogeneously stoichiometric samples are a necessity. In addition, it is important to carry out measurements on the same batch of samples in order to develop a unified picture for a single doping level. In this paper we report magnetic susceptibility, specific heat capacity and transport measurements and investigate the nature of the SDW and sodium ordering transitions in $\rm Na_xCoO_2$, focussing on single crystal samples with x = 0.71.

\section{Experimental details}
The $\rm Na_xCoO_2$ materials were first synthesized in polycrystalline form by means of a solid state reaction of $\rm Na_2CO_3$ and $\rm Co_3O_4$ mixed in a sodium to cobalt ratio of 0.75:1.00. The powders underwent calcination and reaction procedures at 750$^\circ$~C for 12 hours and 850$^\circ$~C for 24 hours respectively, with intermediate grindings to ensure homogeneity. Due to the high volatility of the sodium, a ``fast heat-up'' method was employed during the preparation of the powder. The powders were then isostatically pressed into rods for crystal growth. Single crystals were grown by the floating-zone technique using a CSI four mirror IR image furnace. Pressures ranging from $2.5$ to $10$ bars of O$_2$ were required to suppress sodium loss and growth speeds of up to $10$ mm/ hr were used. 
\par
Crystal boules of 5 mm diameter and up to 70 mm in length were produced. X-ray diffraction of the polycrystalline powder as well as a ground sample of a small portion of the single crystal revealed that the materials were single phase (down to a resolution level of 3\%) hexagonal structures (space group P6$_3$/mmc) with lattice parameters \textit{a} = 2.83 $\rm \AA$ and \textit{c} = 10.97 $\rm \AA$. X-ray Laue images of the cleaved crystals confirmed that the \textit{c}-axis lies perpendicular to the cleavage plane. Despite the precautions taken against sodium loss, a white powder, identified as Na$_2$O$_3$ by X-ray diffraction analysis~\cite{xray}, was observed inside the quartz tube of the image furnace implying a reduction in sodium concentration. Flame photometry analysis of an as-grown, nominally $x=0.75$ crystal revealed a loss of 5.3\% Na during the crystal growth to a reduced stoichiometry of $x=0.71$. All the measurements presented below are taken on samples from this crystal.
\par
Magnetic susceptibility (M/H) measurements were carried out using a Quantum Design MPMS-5S superconducting quantum interference device (SQUID) magnetometer and an Oxford Instruments vibrating sample magnetometer (VSM); AC susceptibility ($\chi_{ac}$) investigations were also carried out using a standard mutual inductance technique with an ac driving field of 1 Oe and a frequency of 403 Hz. Heat capacity and transport data were collected with a Quantum Design Physical Properties Measurement System (PPMS) in magnetic fields up to 70 kOe. The heat capacity (C) measurements were carried out by a two-tau relaxation method using both a standard $^4$He and a $^3$He insert. A background signal (platform and Apiezon N grease) was recorded versus temperature for each run. The dc resistivity ($\rho$), and magnetoresistance (MR) measurements were made using current densities of 0.02-0.2 A/cm$^2$. Fine silver wires were attached to the crystals using silver epoxy. A standard four-probe geometry was used for the measurements of $\rho$ in the \textit{ab} plane. Measurements along the \textit{c}-axis were made with the current (I) contacts attached to the \textit{ab} planes of the crystals and the voltage leads either placed in line along the \textit{c}-axis or on the \textit{ab} faces in a Montgomery configuration.  

\section{Results and Discussion}

\subsection{Magnetic Susceptibility}
\begin{figure}[h]
\centering
\includegraphics[width=10cm]{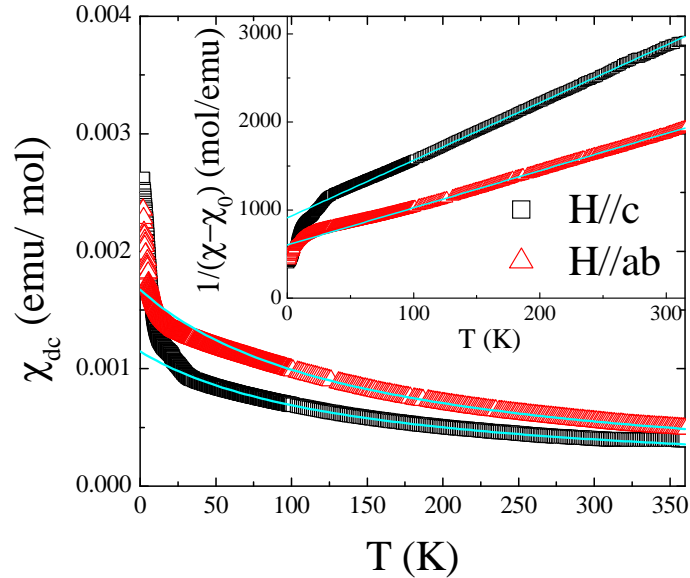}
\caption{DC susceptibility of $\rm Na_{0.71}CoO_2$ measured in an applied field of 1 kOe. The data is anisotropic with $\chi_{\textit{ab}}$ = $\frac{4}{3}$ $\chi_{\textit{c}}$ and exhibits Curie-Weiss behaviour above $\sim$60 K (fitted solid lines). Inverse $\chi_{dc}$ with the constant van Vleck term removed is shown in the inset. The anisotropic gradients are indicative of the anisotropic nature of the g-factor.}
\label{dcsus}
\end{figure}

DC susceptibility ($\chi_{dc}$) was measured between temperatures of 1.8 K and 400 K (see figure~\ref{dcsus}). $\chi_{dc}$ is anisotropic with the measured value for H//\textit{ab} typically $\frac{4}{3}$ times larger than for H//\textit{c} and a changeover in anisotropy at $\sim$6 K. The data can be fitted to a Curie-Weiss term plus a constant $\rm \chi_0$ of $\rm 5\times10^{-5}$ emu/ mol for H$\parallel$ \textit{c} and $\rm 1\times10^{-5}$ emu/ mol for H$\parallel$\textit{ab}. This constant comprises of a small ($\sim$ $\rm 4\times10^{-5}$ emu/ mol) diamagnetic contribution from the localized core electrons and a significant contribution from orbital paramagnetism analogous to van Vleck paramagnetic susceptibility plus a Pauli paramagnetic component arising from the s electrons. After subtraction of $\rm \chi_0$, the inverse susceptibilities (inset of figure~\ref{dcsus}) are linear and, assuming an isotropic free electron value of the Land\'e g-factor, yield values for the Co$^{4+}$ effective moment of 2.05$\rm \mu_B$ and 2.56$\rm \mu_B$ for H$\parallel$c and H$\bot$c respectively. The anisotropy in the measured moments is enigmatic but has been reported by other authors~\cite{wang, chou2}; the same inequitable moments of 1.46$\rm \mu_B/Co^{4+}$ and 2.27$\rm \mu_B/Co^{4+}$ were also observed in one of our $\rm Na_{0.63}CoO_2$ crystals. Chou et al.~\cite{chou2} rationalise their similar results with an anisotropic Land\'e g-factor.
\par
In the best quality samples the data is non-hysteretic; high temperature hysteresis observed by other authors~\cite{prabhakaran} and also in our samples that have been shown to contain cobalt oxide impurities by powder x-ray diffraction, is therefore not considered to be intrinsic to $\rm Na_xCoO_2$ but is attributed to magnetic impurities.
\par
\begin{figure}[h]
\centering
\includegraphics[width=16cm]{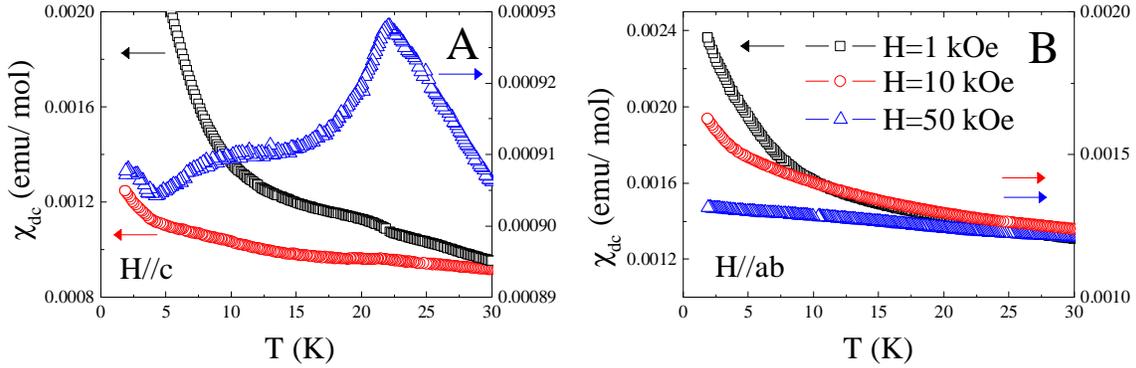}
\caption{Low temperature dc susceptibility versus temperature data for $\rm Na_{0.71}CoO_2$. Panel A: The SDW transition is clearly visible in low fields as a step in $\chi_{dc}$ and as a cusp in higher fields with the field oriented along \textit{c} which corresponds to the AFM spin ordering with the moments aligned along \textit{c}. Panel B: $\chi_{dc}$ data collected over the same temperature range with the field oriented along the \textit{ab} planes. No features that can be associated with the SDW are observed.}
\label{SDW}
\end{figure}

In low fields at T$_{mag}$=22 K, a shoulder in the magnetisation is visible in data measured along the \textit{c}-axis (figure~\ref{SDW} panel A) with no corresponding feature discernable in the \textit{ab} plane (figure~\ref{SDW} panel B). The susceptibility data can be explained as the establishment of antiferromagnetic spin/charge density wave within the cobalt oxide layers whose moments are aligned antiferromagnetically along \textit{c}; in contrast to other published data~\cite{sales} there is no evidence for a canting of the moments into the \textit{ab} plane. A similar behaviour has also been observed in the related layered cobaltate $\rm Ca_3Co_4O_9$, where the spin density wave can only be distinguished in the \textit{c} plane susceptibility~\cite{sugiyama3}. The step in $\chi$ at T$_{mag}$ can be observed even in very low fields of 10 Oe; with an increase in applied magnetic field the step becomes more distinct. At the highest fields the transition is qualitatively different and more reminiscent of an antiferromagnetic cusp, accentuating the fact that the principal magnetic interaction in this system is antiferromagnetic in nature. Because of the lack of any significant impurity levels in our samples, the upturn in $\chi$ at low T cannot be attributed to magnetic impurities. The magnitude of the Curie constant in the low temperature data corresponds to the contribution of all of the Co$^{4+}$ spins in the sample; we can therefore rule out the possibility of a Curie tail arising from free spins at the ends of broken spin chains. Although the upturn in suppressed at high fields, this cannot be due to saturation of free moments as we have found that M(H) does not saturate even in fields as high as 120 kOe. The transition temperature is independent of applied field and does not display hysteretic behaviour indicating that this is a second order phase transition. 
\par
\par
\begin{figure}[h]
\centering
\includegraphics[width = 10 cm]{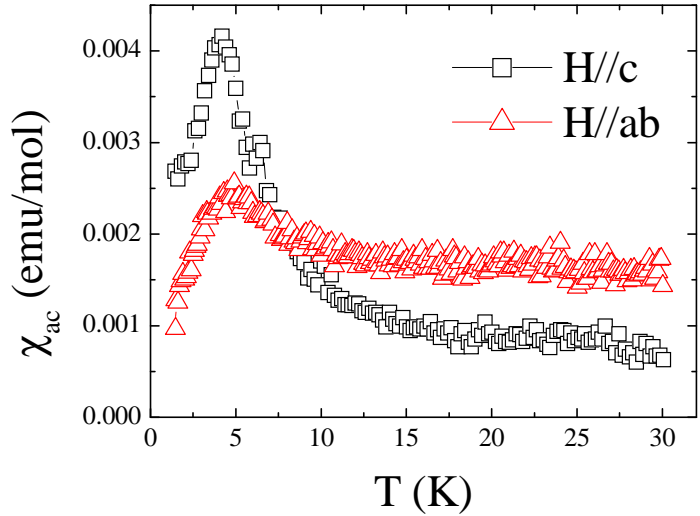}
\caption{Low temperature ac susceptibility of $\rm Na_{0.71}CoO_2$ versus temperature. No feature associated with the spin density wave transition at 22K is visible, however, the cusp at 4 K indicates that the break up of the SDW has resulted in a glassy ground state.}
\label{acsus}
\end{figure}

Low temperature ac susceptibility measurements were taken with the driving field oriented along the \textit{c}-axis and in the \textit{ab} plane and are shown in figure~\ref{acsus}. $\chi_{ac}$ displays the same anisotropy and low temperature cross-over as $\chi_{dc}$. Curiously, no magnetic features were visible at T$_{mag}$. A peak at $4$ K is visible for fields applied along both \textit{ab} and \textit{c}, which has previously been reported~\cite{shi} and attributed to low temperature antiferromagnetic ordering. This low temperature feature is reminiscent of the susceptibility cusps found in spin glasses. Preliminary investigations of the frequency dependence show a decrease in the cusp temperature with decreasing frequency, which is consistent with the varying dynamic response expected from a glassy ground state. The origin of this state is likely to be a consequence of the incoherency of the SDW with the periodicity of the lattice. It is plausible that the density wave may become pinned to the underlying Co lattice which would result in low temperature 'glassy' behaviour. Other materials in which a spin density wave state has been observed, for example the organic salt $\rm (TMTSF_2)PF_6$~\cite{biljakovic}, also exhibit low temperature glassy states for this very reason.
\par

\par
\begin{figure}[h]
\centering
\includegraphics[width = 16 cm]{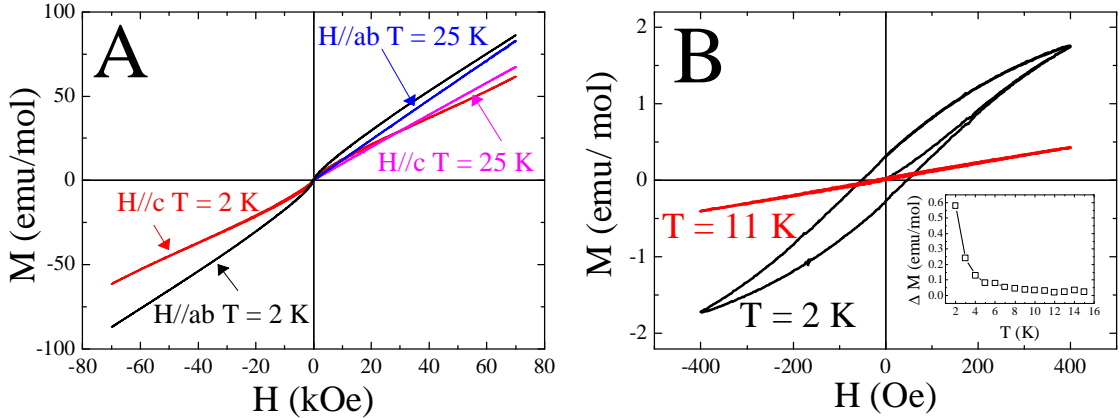}
\caption{Panel A: Magnetisation loops measured as a function of field for $\rm Na_{0.71}CoO_2$ with the sample oriented with H$\parallel$\textit{c} and H$\parallel$\textit{ab} at two different temperatures. Panel B: Low field hysteresis as measured with H$\parallel$\textit{c}. The remanent magnetisation clearly visible at 2 K has almost disappeared in the higher temperature (11 K) data; the magnitude of this ferromagnetic component is plotted as a function of temperature in the inset.}
\label{MHloops}
\end{figure}

Magnetisation data as a function of applied field were collected above and below T$_{mag}$ in order to further characterise the nature of the SDW and are shown in figure~\ref{MHloops}. Once again, $\rm M_{\textit{ab}} > M_{\textit{c}}$ (panel A) and the anisotropy in the highest measured fields is independent of temperature at $\sim$1.4, which is in line with the $\rm \chi_{dc}$ vs. T results. At the lowest temperatures the loops exhibit considerable curvature (shown in panel B), which disappears above 15 K. At 2 K the value of the magnetisation at 70 kOe corresponds to moments of only $\sim$0.016 $\mu_{B}$/Co and $\sim$0.011 $\mu_{B}$/Co along \textit{ab} and \textit{c} respectively. Above 60 kOe there is a slight increase in $\partial M_{c}/\partial H$, although no spin-flop transition is observed up to fields as high as 120 kOe~\cite{sales}. At low fields the data has hysteretic behaviour, which disappears at a temperature of $\sim$15 K (inset of panel B). The remanent magnetisation at 2 K is extremely small corresponding to just 0.0001 $\mu_{B}$/Co along \textit{c} and an order of magnitude smaller along \textit{ab}. The disappearance of this low temperature ferromagnetic component has been noted by previous authors~\cite{motohashi}, although it was reported to occur below T$_{mag}$, whereas our data indicate a lower temperature transition. Anomalies in magnetic susceptibility data below $T_{mag}$ have also been observed by Sakurai et al.~\cite{sakurai} and attributed to modifications to the magnetic ground state. Again, our data are consistent with the appearance, immediately below $T_{mag}$, of a SDW with the magnetic moments aligned antiferromagnetically along \textit{c}, with the appearance of a weak ferromagnetic component below 15 K. The co-existence of a ferromagnetic ground state with a SDW state (both with different ordering temperatures) has previously been seen in the related material $\rm Ca_3Co_4O_9$~\cite{sugiyama3}, as well as several other SDW wave systems e.g. $\rm UNi_2Si_2$~\cite{lin}. In these materials the small ferromagnetic component arises from magnetic anisotropy, modifications to the geometry of the Fermi surfaces as function of temperature below $T_{mag}$, and from the structure which leads to differences between the inter- and intra-layer coupling.

\subsection{Specific Heat Capacity}

\begin{figure}[h]
\centering
\includegraphics[width = 10cm]{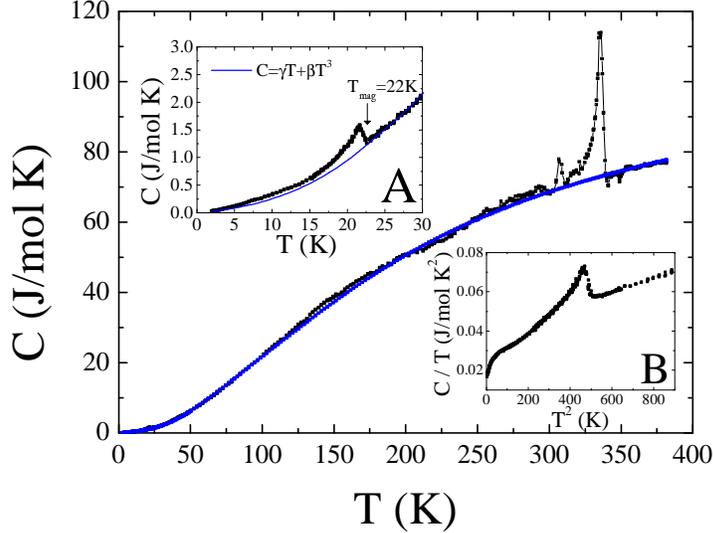}
\caption{Specific heat capacity data for $\rm Na_{0.71}CoO_2$. Three transitions are visible; two high temperature peaks at 307 K and 336 K are attributed to sodium ordering within the structure and the SDW is visible as a peak at 22 K (panel A). The solid line in the main figure corresponds to the combined Debye/Einstein model; the line in panel A is a fit to $\gamma$T + $\beta$T$^3$ above T$_{mag}$. Low temperature C/T vs. T$^2$ data (panel B) shows a hump centered at 7 K, which validates the notion of a low T glassy state.}
\label{cdata}
\end{figure}

We have measured the heat capacity of $\rm Na_{0.71}CoO_2$ single crystals from 400 mK to 380 K (see figure ~\ref{cdata}). Over the whole temperature range studied the data cannot be fitted using a single Debye expression. Satisfactory agreement with the data can be obtained using a combined Debye - Einstein function giving $\Theta_D$ of 410 K and $\Theta_E$ of 820 K weighted in the ratio 4/5. This appears to be consistent with a mixture of acoustic and optical modes expected from a combination of light and heavy elements.
\par
The high temperature data contains two distinct features, one at T$_1$=307 K and a second much larger peak at T$_2$=336 K. We estimate from x-ray and magnetisation data that the magnetic impurity levels of $\rm Co_3O_4$ or CoO in the sample are less than 3\%. $\rm Co_3O_4$ orders antiferromagnetically at 40 K~\cite{LB}; there are no features in the data at this temperature. CoO is reported to be AFM at T$_N$=289 K~\cite{zheng}. In order to rule out effects from CoO  we have measured the heat capacity of a polycrystalline sample uniaxially pressed into the form of a thin (3 $\times$ 3 $\times$ 1.5 mm$^3$) pellet~\cite{aldrich}; a clear feature is seen in the data at T$_N$=289 K. It maybe that non stoichiometric CoO is present in the sample or that CoO is incorporated into the layered structure altering T$_N$ and the thermodynamic response. While we cannot unequivocally rule out impurities producing these features we conclude at this stage that the high temperature peaks seen in the $C(T)$ curves are intrinsic to the sample. We attribute the transition at T$_2$ to a real space ordering of the Na within the system. This sodium ordering transition has been noted by previous authors~\cite{huang}, in which the sodium is rearranged from random ordering on the $\rm 6h(2x,x,\frac{1}{4})$ to the higher symmetry $\rm 2c(\frac{2}{3},\frac{1}{3},\frac{1}{4})$ site at sodium concentrations above x=0.75. The entropy associated with this transition is $\sim$1 J/ mol K. This compares to just 10\% of the configurational entropy expected for sodium ordering over the two available sites in the correct ratios, which is calculated to be 1.2R. This may indicate that considerable Na disorder persists below T$_2$.
\par
Between a limited temperature range of 22 K to 30 K the data can be explained by assuming that C comprises of two terms, an electronic term and a phonon component, which in the low temperature limit, can be characterised using the Debye model giving C=$\gamma$T+$\beta$T$^3$. A linear fit to the data (inset B of figure~\ref{cdata}) gives a $\gamma$ of 35 mJ/ mol K and $\Theta$$_D$=550 K. The value of $\Theta_D$ we obtain is slightly higher than the values calculated using the high temperature data but is in line with previous estimates. Using this value we find that above 40 K the experimental data deviates substantially from the Debye theory. This is consistent with a smaller than expected value for the measured heat capacity at high temperature. We find [C(380 K)/ (3.7$\times$3R)]$_{measured}$=0.7 while the expected ratio using the Debye theory is 0.9. 
\par
There is a lambda like anomaly at 22 K (inset A of figure~\ref{cdata}) indicating the onset of magnetic ordering. This feature has previously been associated with the development of a SDW within this material and our results are found to be consistent with this interpretation. The transition is rather broad extending over at least 7 K, whereas the minimum sample heat pulse used around the transition was 0.1 K. This feature occurs at the same temperature in both heating and cooling runs with no discernable hysteresis to within the experimental accuracy of the technique used, again indicating that this is a second-order phase transition~\cite{pulse}. The jump at T$_{mag}$ is 0.4 J/ mol K which corresponds to 25\% of the signal at this temperature. Using the standard BCS expression $\Delta$C/T$_{SDW}$=1.43$\gamma$ gives $\gamma$ =12.7 mJ/ mol K which is consistent with the values obtained from the low temperature heat capacity (see below). This value is large compared to the value seen in the spin density wave compound $\rm (TMTSF)_2PF_6$, where the jump in the specific heat at the SDW transition (T=12.1 K) amounts to only $1.5\%$ of the total heat capacity,~\cite{odin, coroneus} but is smaller than the $134\%$ increase seen in $\rm URu_2Si_2$ at 17.5 K~\cite{maple}.
\par
Below T$_{mag}$ there is a considerable reduction in $\gamma$. An extrapolation of the low temperature data produces a value of 15 mJ/ mol K. These values agree well Bayrakci et al.~\cite{bayrakci} who calculated $\gamma$ for an x=0.82 sample from data below 22 K. It is interesting that the $\gamma$ value below the SDW transition also agrees with estimates for $\gamma$ made for the superconducting material $\rm Na_{0.3}CoO_2\bullet1.3H_2O$ ~\cite{cao, jin, chou}. Assuming the $\gamma$ term arises solely from charge carriers we can use $\gamma$=($\pi^2/2)$$k^2_B$N(E$_F$) to calculate the free electron density of states. N(E$_F$) is calculated to be 5.96$\times$10$^2$4 eV/ mol just above T$_{mag}$ reducing to 2.55$\times$10$^{24}$ eV/ mol at 2 K. The ratio of $\gamma$ above and below the transition suggests that $50-60\%$ of the Fermi surface is removed by the opening up of a gap at T$_{mag}$.
\par
\par
The entropy associated with this feature can be obtained by subtracting from the total specific heat, C, an estimate of the background specific heat, C', made up of a contribution from the ungapped electrons plus the phonon contribution which we assume remains unchanged between base and 30 K (see figure~\ref{cdata}). The entropy associated with the anomaly alone (i.e. from T=14 to 25 K) amounts to 0.085 J/ mol K and corresponds to only 5\% ($\frac{3}{10}$Rln($2S+1$)=1.73 J/ mol K) of the entropy expected for a $x=0.71$ sample with a Co$^{3+}$ (spin 0) / Co$^{4+}$ (spin $\frac{1}{2}$) system in the ratio of 3:1. Including the excess entropy down to 2 K this value rises to 0.16 J/ mol K. The value compares well with estimates made in other reports~\cite{bayrakci, motohashi} and is consistent with previous reports for SDW compounds. In the case of Cr~\cite{overhauser}, the entropy at the SDW transition (0.0184 J/ mol K) corresponds to only $0.6\%$ of the entropy associated with long range antiferromagnetic order, but is equivalent to $4\%$ of the total electronic entropy, while in the case of $\rm (TMTSF)_2PF_6$, the entropy associated with the heat capacity anomaly amounts to $10\%$ of the total electronic entropy ~\cite{coroneus}.  In our case a low temperature $\gamma$ of 15 mJ/ mol K suggests that the total Sommerfeld electronic entropy at T=22 K should be 0.33 J/ mol K. The entropy associated with the transition is approximately $50\%$ of this value. We can model $\delta$C=C-C' using $\delta$C=$A\exp(-\Delta/T)$ between 15 and 22 K and obtain an estimate for $\Delta$=163 K (11.2 meV) = 7.4T$_{mag}$~\cite{maple}. This value is much larger than the weak coupling BCS value and indicates that a  strong electron-phonon coupling is involved in the SDW transition. Similar values have been seen in the case of $\rm URu_2Si_2$~\cite{maple} and $\rm TaSe_2$, $\rm TaS_2$, and $\rm NbSe_2$ ~\cite{wang3}. It would be interesting to compare the value of $\Delta$ obtained here with estimates made using other techniques, for example, inelastic neutron diffraction.
\par
There also exists an additional contribution to the heat capacity at 7K, which is clearly visible in the C/T vs. T$^2$ plot as a broad (extending from 4 - 15 K) bump (inset B of figure~\ref{cdata}). The low temperature nodule is consistent with the notion of a glassy state, which was acknowledged in section $3.1$. The broad peak is a result of the pinning of SDW domains to over an extended temperature range and has been observed for previously studied spin density wave systems~\cite{biljakovic}.
\par
Below 4 K the C versus T dependence can once again be described by $\rm C = \gamma T + \beta T^3$. In contrast to the work of both Br\"{u}hwiler et al.~\cite{bruhwiler} who studied an x=0.7 sample and Ando et al.~\cite{ando} who have reported on heat capacity data for $\rm NaCo_2O_4$ we observe no upturn in C/T at low temperature. It is not clear if the upturn in C/T has the same origin for the two cases that are reported. It is apparent that the SDW transition is absent in the material studied in reference~\cite{ando}, while the C versus T data around 22 K is not shown in reference~\cite{bruhwiler}. It would be instructive to study a range of Na doping levels in order to determine if the more highly correlated electron behaviour reported in x=0.5 is suppressed as x is increased.

\subsection{Transport Data}

\begin{figure}[h]
\centering
\includegraphics[width = 16 cm]{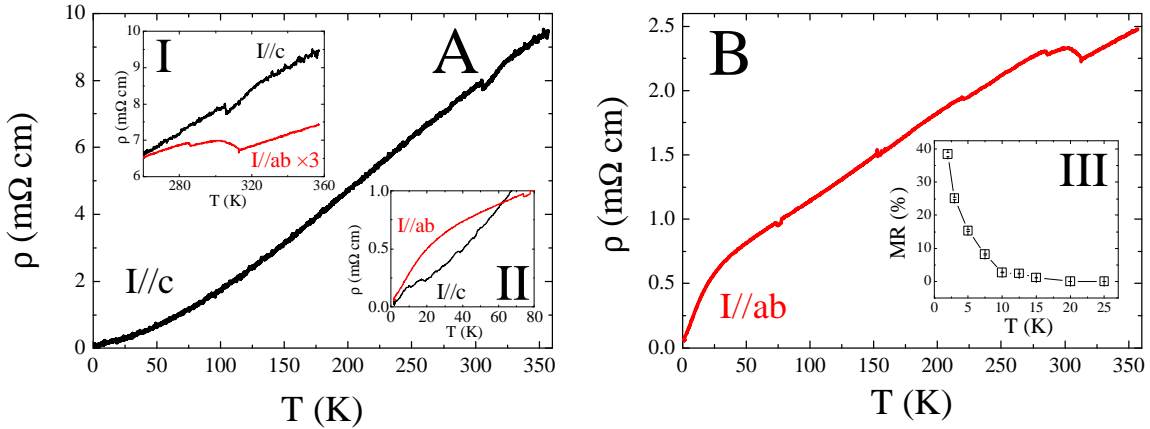}
\caption{Zero field resistivity data with the current oriented along the \textit{c} axis (panel A) and \textit{ab} planes (panel B). Inset I shows the high temperature sodium ordering transitions for both curves, whilst inset II shows details of the low temperature features, namely the cross-over in anisotropy and increase in $\partial \rho/\partial T$ along \textit{ab}. Positive magnetoresistance as a function of temperature as measured with an applied field of 70 kOe with the current oriented along \textit{ab} is shown in inset III; no corresponding magnetoresistance was observed along \textit{c}.}
\label{tdata}
\end{figure} 
\par

The values of the resistivity at room temperature in both the \textit{c} (figure~\ref{tdata} panel A) and \textit{ab} (figure~\ref{tdata} panel B) directions are of the order 1-10 m$\Omega$ cm. These values are in reasonable agreement with several other reports of resistivity measurements made on either polycrystalline or single crystals samples of $\rm Na_{0.7}CoO_2$ in the \textit{ab} plane ~\cite{wang, foo, motohashi, bayrakci, wang2}. At room temperature the resistivity is only slightly anisotropic with $\rho{_\textit{c}}$ $\approx$ 3$\rho{_\textit{ab}}$ and a crossover in anisotropy at $\sim$60 K (figure~\ref{tdata} inset II). The resistance ratios (R350K/R2K) along \textit{ab} and \textit{c} are 30 and 200 respectively; at 2 K $\rho{_\textit{ab}}$ and $\rho{_\textit{c}}$ take values of 80 and 45 $\mu\Omega$ cm respectively. The data for $\rm Na_{0.71}CoO_2$ reported here contrast sharply with previous publications for the x=0.5 composition which report an anisotropy ratio $\rho{_\textit{c}}$ / $\rho{_\textit{ab}}$ as high as 200 at 4.2 K and a maximum in $\rho{_\textit{c}}$ at $\approx$180 K resulting from a crossover from insulating two dimensional to metallic three dimensional behaviour ~\cite{terasaki, valla}. We find that in both directions the resistivity between 280 K and 100 K varies almost linearly with temperature with d$\rho$/dT=6.4 $\mu\Omega$ cm/ K and 30.0 $\mu\Omega$ cm/ K for the \textit{ab} and \textit{c} directions respectively. A significant reduction in $\rho{_\textit{ab}}$ below 40 K (with a maximum in d$\rho{_\textit{ab}}$/dT at 10 K) marks the SDW transition. No corresponding feature is seen in $\rho{_\textit{c}}$. The reduction in $\rho$ suggests that the opening up of a gap associated with the SDW leads to a decrease in the total scattering within the ab plane while the conduction along \textit{c} is unaffected. This downward feature is observed in several of the $\rho$ versus T curves for $\rm Na_xCoO_2$ ($x\approx0.7$) reported in the literature ~\cite{foo, motohashi}. It is missing in other reports~\cite{wang2}, although notably the spin density wave transition is also absent in the susceptibility data. Two previous groups~\cite{bayrakci, mikami} have reported an increase in $\rho$ at T$_{mag}$ at low temperatures albeit on higher sodium doped samples (x = 0.82 and 0.9 respectively). However, in the most recent report~\cite{sakurai2} for a range of x between 0.7 and 0.78, which includes the composition studied here, no upturn is observed. 
\par
At high temperature for $\rho{_\textit{ab}}$ there are two weakly hysteretic transitions at 290 K and 320 K. For $\rho{_\textit{c}}$ there is a single feature at 309 K (figure~\ref{tdata} inset I). There is also a $\sim$20\% decrease in d$\rho$/dT at higher temperatures (e.g. the value of d$\rho{_\textit{ab}}$/dT decreases from 6.42 $\mu\Omega$ cm/ K to 7.1 $\mu\Omega$ cm/ K above 315 K). We associate these higher T features with an order-disorder transition of the Na ions (see heat capacity). Refinement of the structure has suggested that at high temperature Na ions are able to migrate through the structure in the planes between the two oxygen sheets~\cite{balsys}. Ionic conduction is thus expected via the direct movement of the sodium ions. The sodium atom can occupy two positions in the lattice and a freezing of the Na ions onto well defined sites could lead to a small increase in $\rho$. Ionic conductivities, however, at room temperature are expected to be much lower (10$^{-3}$ S/ cm) than the magnitude of the jump in $\rho$ seen around room temperature ($\Delta\rho$= 0.1 m$\Omega$ cm). Foo et al.~\cite{foo} have suggested that there are strong correlations between the Na ions and the charge carriers (in this case spin-$\frac{1}{2}$ holes ($Co^{4+}$) hopping in a diamagnetic background of $Co^{3+}$ ions). At x=$\frac{1}{2}$ this leads to an insulating state at low T. At x=0.71 the effects are expected to be weaker. Nevertheless, the carriers are still likely to be influenced by modulations in the Na layers as the sodium ions order leading to charge localisation. These effects are much more strongly felt within the $\rm CoO_2$ layers resulting in the sharp increase in $\rho$.

\par
Positive magnetoresistance has been observed for the in-plane data (figure~\ref{tdata}, inset III). No magnetoresistance was seen for $\rho_\textit{c}$. Changing the sample set up from H$\parallel$I to H$\bot$I did not produce any noticeable changes. At a temperature of 2 K and a field of 70 kOe the resistivity was seen to increase by 40$\%$, consistent with other published data~\cite{motohashi}. The magnetoresistance obeys Kohler's rule (with no indication of saturation up the maximum measured field of 70 kOe) and is removed by raising the temperature above $\sim$15 K. It is interesting to note that this correlates well with the temperature at which the spontaneous magnetisation present in the $\chi$(H) data disappears.
\par
Electronic structure calculations for $\rm NaCo_2O_4$~\cite{singh} give a resistivity anisotropy ($\rho{_\textit{c}}$ / $\rho{_\textit{ab}}$) of 8. The higher values seen experimentally for x=0.5~\cite{valla} are attributed to the effects of scattering due to Na disorder, which it is claimed should make the $\textit{c}$-axis transport less coherent. The increased Na content in these samples appears to lead to a significant modification in the transport process producing a more three dimensional system. For $\rm Na_xCoO_2$ the Co-Co distance remains almost constant at 2.83 $\rm \AA$ for $0.3 \leq x \leq 0.75$. We note that this Co-Co spacing is above the critical value quoted by Molenda~\cite{molenda} for metallic conduction. The \textit{c}-axis decreases with increasing x. At x=0.5 a crossover from metallic behaviour at low temperature to thermally activated polaronic conduction at room temperature has been attributed to an increase in the \textit{c}-axis distance with temperature~\cite{rivadulla}. This in turn leads to a transition from itinerant to a more localised ${\textit{a}}{_\textit{1g}}$ band as the temperature increases. Following the arguments of Molenda we suggest that at x=0.7 the shorter out of plane Co-O-O-Co bond length (here the room temperature value of \textit{c} is 10.96 $\rm \AA$) stabilises the itinerant electron phase along the c-axis while the extended ${\textit{a}}{_\textit{1g}}$ $+$ ${\textit{e}}{\textit{'}}{_\textit{g}}$ band ensures good conductivity within the \textit{ab} plane. In addition the presence of increased Na content reduces the tendency for carrier localisation along \textit{c} because of the preferential random ordering of the sodium on the 6h site as mentioned previously~\cite{huang}.

\section{Conclusions}
We have identified a number of features in our data that are expected to be common to high quality single crystal samples in the x$\sim$0.7 region of the phase diagram. An antiferromagnetic SDW ground state exists below 22 K. The onset of the SDW has effects on the heat capacity and transport properties, causing a reduction in both $\gamma$ and $\rho$ below T$_{mag}$. It has been suggested for $\rm Na_xCoO_2$ that the two sections of the Fermi surface (i.e. the narrow ${\textit{a}}{_\textit{1g}}$ and the broader ${\textit{a}}{_\textit{1g}}$ $+$ ${\textit{e}}{\textit{'}}{_\textit{g}}$) bands play different roles. The magnetic susceptibility, $\chi$, and the electronic component of the heat capacity, $\gamma$, depend (to lowest order) on the density of states which is determined by the large Fermi surface with ${\textit{a}}{_\textit{1g}}$ symmetry. The carriers in the ${\textit{a}}{_\textit{1g}}$ $+$ ${\textit{e}}{\textit{'}}{_\textit{g}}$ band are mobile because the band is spread in the \textit{ab} plane. The ${\textit{a}}{_\textit{1g}}$ band is unstable to the formation of the SDW leading to modifications in $\chi$ and $\gamma$. The reduction in electron-electron scattering resulting from the formation of the SDW leads to an increase in conductivity along \textit{ab}, despite the expected reduction in the number of charge carriers.
\par
The layered structure of the $\rm Na_xCoO_2$ system is reflected in the magnetic susceptibility data where both the temperature independent susceptibility and Land\'e g-factor are anisotropic, a feature common to other cobalt based oxide materials. The transport properties are also anisotropic. This anisotropy appears to increase with decreasing x (towards x = $\frac{1}{2}$). Other features also appear to depend sensitively on doping levels. For instance, the Na ordering transition has been reported to occur at various different temperatures in the range 320-340 K. This may be a consequence of the fact that the relative occupancies of Na(1) and Na(2) are variable with x.
\par
Below 15 K the hysteresis observed in M(H) loops indicates that the magnetic order has a small ferromagnetic component. A large, positive magnetoresistance coexists with this FM state. The magnitude of the observed hysteresis along with the nearly linear M(H) behaviour and the peak in $\chi_{dc}$(T) seen at higher fields, underline the fact that whilst ferromagnetic correlations play a role in this system, the SDW is predominantly antiferromagnetic in nature. At the lowest temperatures both the heat capacity and the $\chi_{ac}$ data suggest that a significant portion of the SDW state is pinned producing a glassy ground state. This is not uncommon amongst SDW materials. 

\ack
We acknowledge the financial support of the EPSRC (UK) for this project.




\begin{thebibliography}{99}
\bibitem{xray} The powder diffraction pattern was identified as $\rm Na_2CO_3$ by comparison to ICDD powder diffraction file 19-1130. No carbonate impurity existed in the crystal precusor material or in the furnace atmosphere and it is therefore assumed that sodium oxide is produced during crystal growth which then reacts with carbon dioxide once the furnace is opened.
\bibitem{pulse} Note that during both warming and cooling runs, the sample is heated by a certain percentage of the set point for data collection.
\end{thebibliography}
\end{document}